\documentclass[letterpaper]{jpconf}
\usepackage{graphicx}
\begin{document}
\title{de Broglie Wave Phase Shifts Induced by Surfaces Closer than \mbox{25 nm}}
\author{Alexander D. Cronin and John D. Perreault}
\address{Department of Physics, University of Arizona,1118 E 4th St,
Tucson, AZ 85721}
\ead{cronin@physics.arizona.edu}
\begin{abstract}
Four atom optics experiments that each serve to measure atom-surface
interactions near nanofabricated gratings are presented here. In
these experiments atoms in a beam travel within 25 nm of a material
grating bar, and the analysis incorporates phase shifts for the
atomic de Broglie waves due to interactions betwen Na atoms and
silicon nitride surfaces. One atom diffraction experiment determines
the van der Waals coefficient $C_3=2.7\pm$0.8 meVnm$^3$, and one
atom interferometer experiment determines $C_3=4\pm$1 meVnm$^3$. The
results of all four experiments are consistent with the Lifshitz
prediction that is explicitly calculated here for Na-silicon nitride
to be $C_3=3.25$ meVnm$^3$. The four atom optics experiments and
review of van der Waals theory are complemented by similar
experiments using electron beams and analysis of image-charge
effects.
\end{abstract}

\section{Introduction}

Ten nm away from a surface the potential energy for an atom is
approximately 3 $\mu$eV, and for an electron it is about 10,000
times larger. More precisely, the van der Waals potential for
sodium atoms and the image-charge potential for electrons both
depend on the permittivity of the surface material; both
potentials are also affected by surface charges, surface coatings,
and surface geometry. Precise knowledge of the potential close to
real surfaces is now needed for understanding atom optics
experiments and nanotechnology devices, yet measurements of
atom-surface interaction strengths have only been made for a few
systems so far. Here we present four atom optics experiments that
serve to measure the potential energy for atoms due to a surface
located within 25 nm. Comparison to theoretical values of the
non-retarded atom-surface van der Waals interaction will be made
in the discussion.

Our experiments are based on coherent transmission of sodium atom
de Broglie waves through an array of 50 nm wide channels in a
silicon nitride nanostructure grating. Transmitted atoms pass
within 25 nm to a grating bar surface, and remain this close for
only $10^{-10}$ sec. Even in this short time, interactions with
the channel walls modify the phase of the atom waves. Phase front
curvature on the nanometer scale has the observed effect of
modifying the phase $\Phi_n$ and amplitude $|A_n|$ in each
far-field diffraction order. We measured atom diffraction
intensities and atom interferometer fringe phase shifts in order
to determine the potential for sodium atoms induced by surfaces of
silicon nitride.

For comparison we used the same nanostructure gratings to diffract
electron beams. Despite the 10,000 times larger potential energy (at
10 nm) due to image-charge effects, electron diffraction shows
similar features to atom diffraction as a result of stronger
interactions with the surface over shorter time scales.

\section{Nanostructure Gratings}

\begin{figure}[h]
\begin{minipage}{18pc}
\includegraphics[width=16pc]{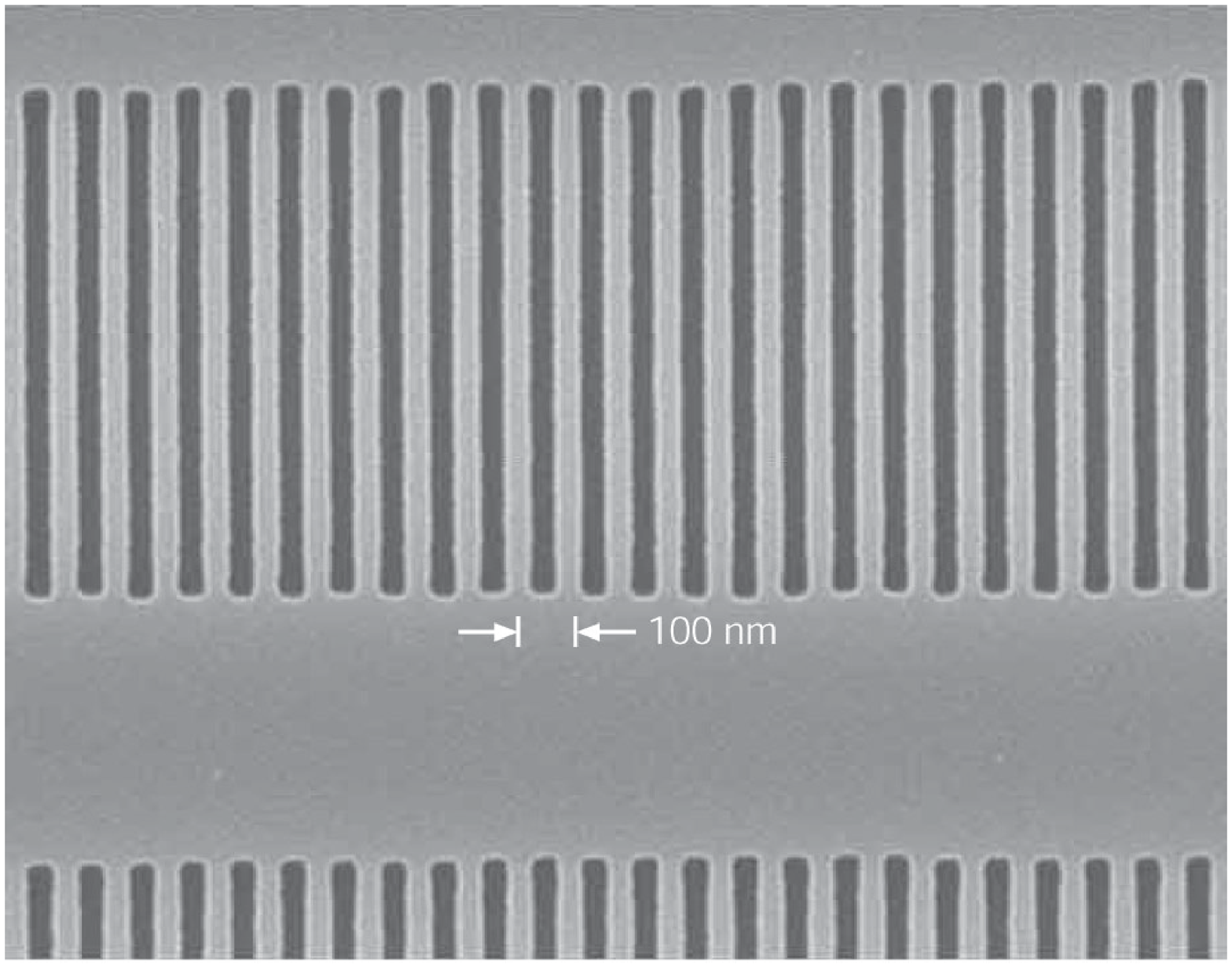}
\caption{\label{g1} Front view of a nanostructure grating with a
100-nm period. The free-standing silicon nitride bars appear light
in this image.  Image courtesy of T.A. Savas at the MIT
NanoStructures Laboratory.}
\end{minipage}\hspace{2pc}%
\begin{minipage}{18pc}
\includegraphics[width=18pc]{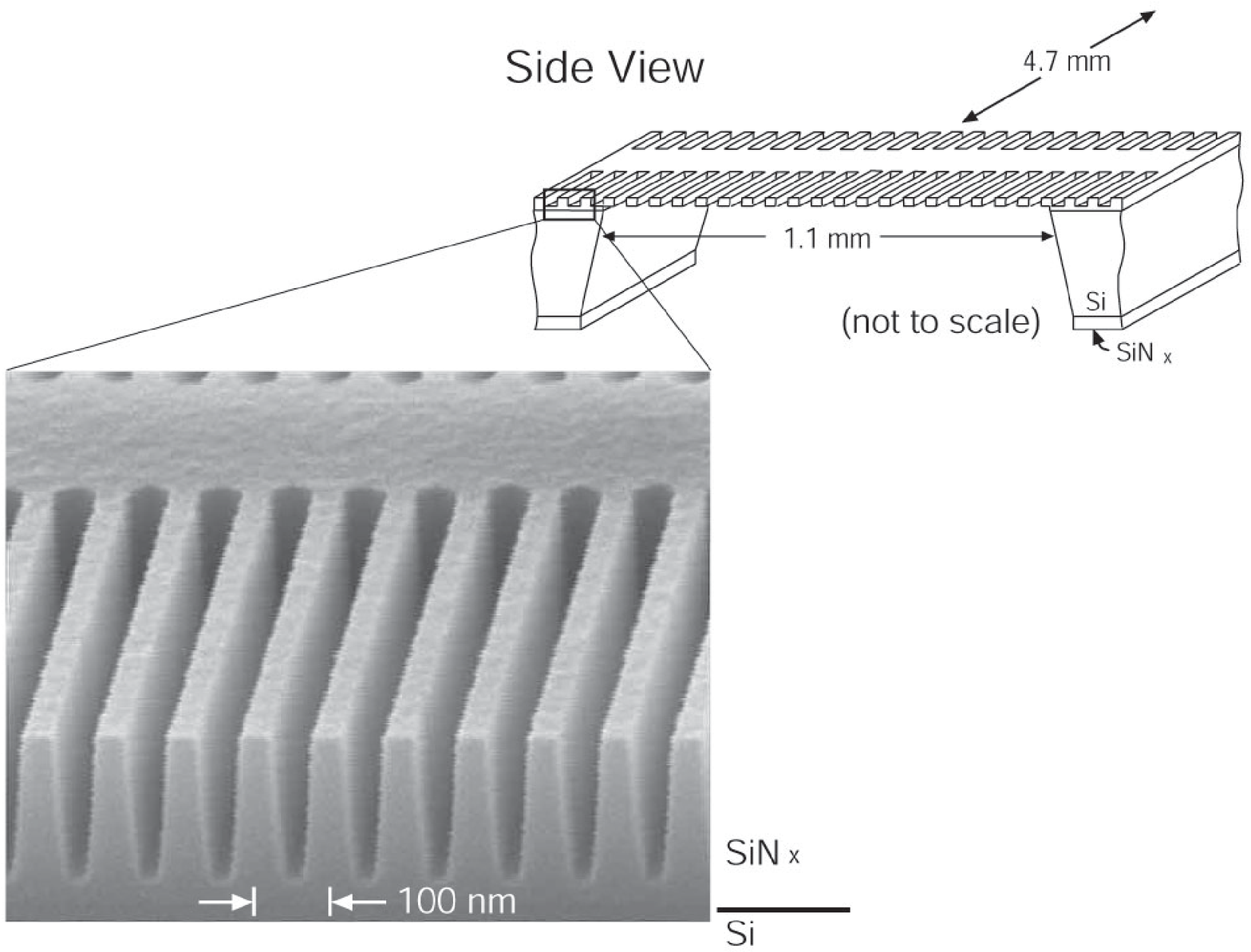}
\caption{\label{g2} Cross-section of a cleaved grating.  Note the
trapezoidal bar profile. Bars in this region are not free standing,
but are attached to a substrate. Image courtesy of T.A. Savas.}
\end{minipage}
\end{figure}

The 100-nm period gratings were fabricated by T.A. Savas at MIT
using photolithography with standing waves of UV laser light. The
etch procedures used to create the silicon nitride nanostructures
are described in \cite{sava96}.  The bars are free-standing
trapezoidal columns and their dimensions have been measured to an
accuracy of 1.5 nm using scanning electron microscope (SEM) images
such as those shown in Figures \ref{g1} and \ref{g2}.

To understand how van der Waals interactions modify the amplitude
and phase of diffracted atom waves, we consider three steps. First,
a model for the potential in all space around the grating bars is
needed. Second, the phase shift and absorption for atom waves
transmitted through this potential is calculated as a function of
position. Finally, the propagation to the far-field is given by a
Fourier transform of the transmission function.

\begin{figure}[h]
\begin{minipage}{18pc}
\includegraphics[width=18pc]{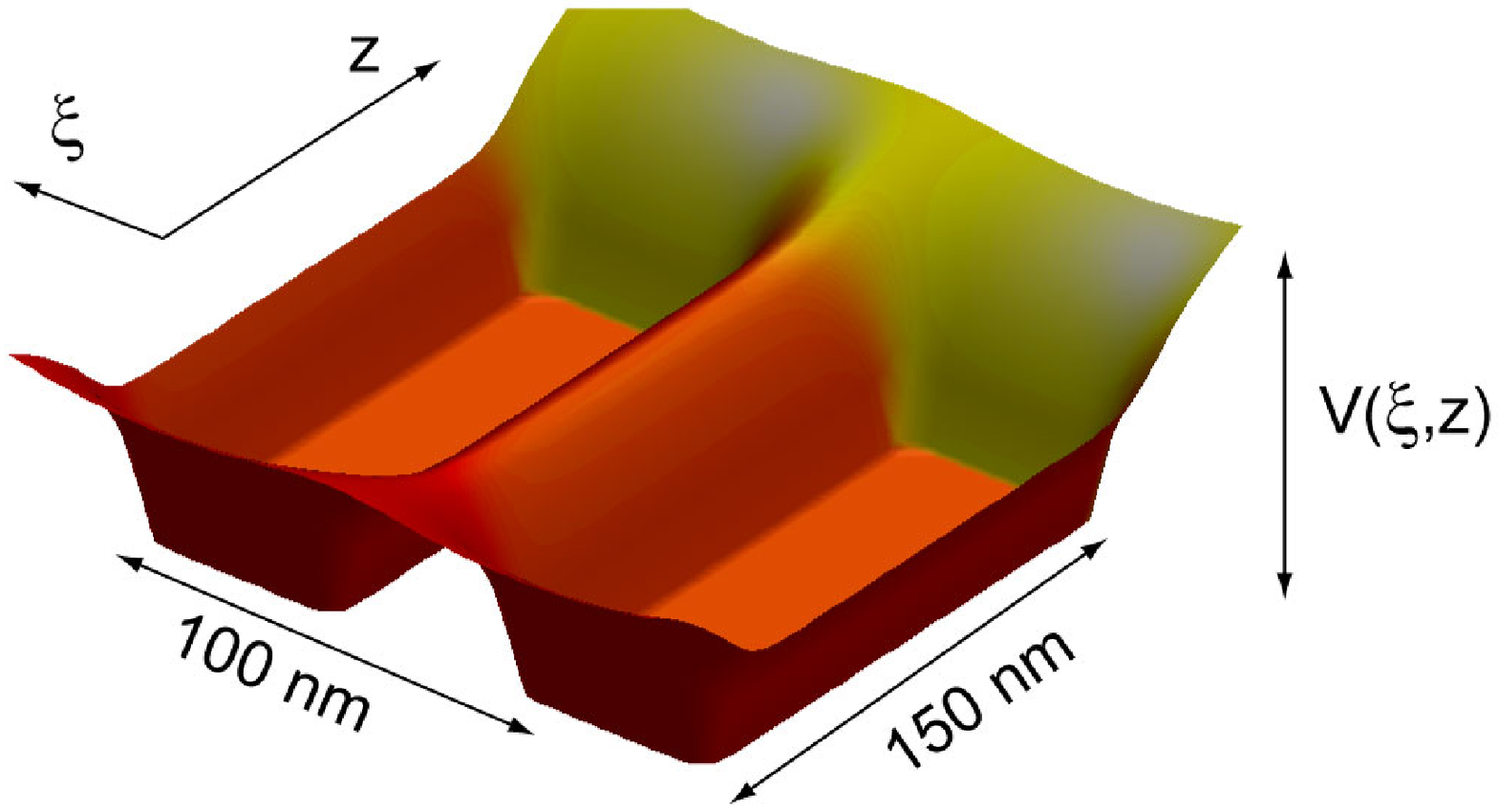}
\caption{\label{3Dpotential} The potential energy $V(\xi,z)$ in the
vicinity of trapezoidal columns can be approximated by a pairwise
interaction. The vertical axis is shown on a log scale.}
\end{minipage}\hspace{2pc}%
\begin{minipage}{18pc}
\includegraphics[width=18pc]{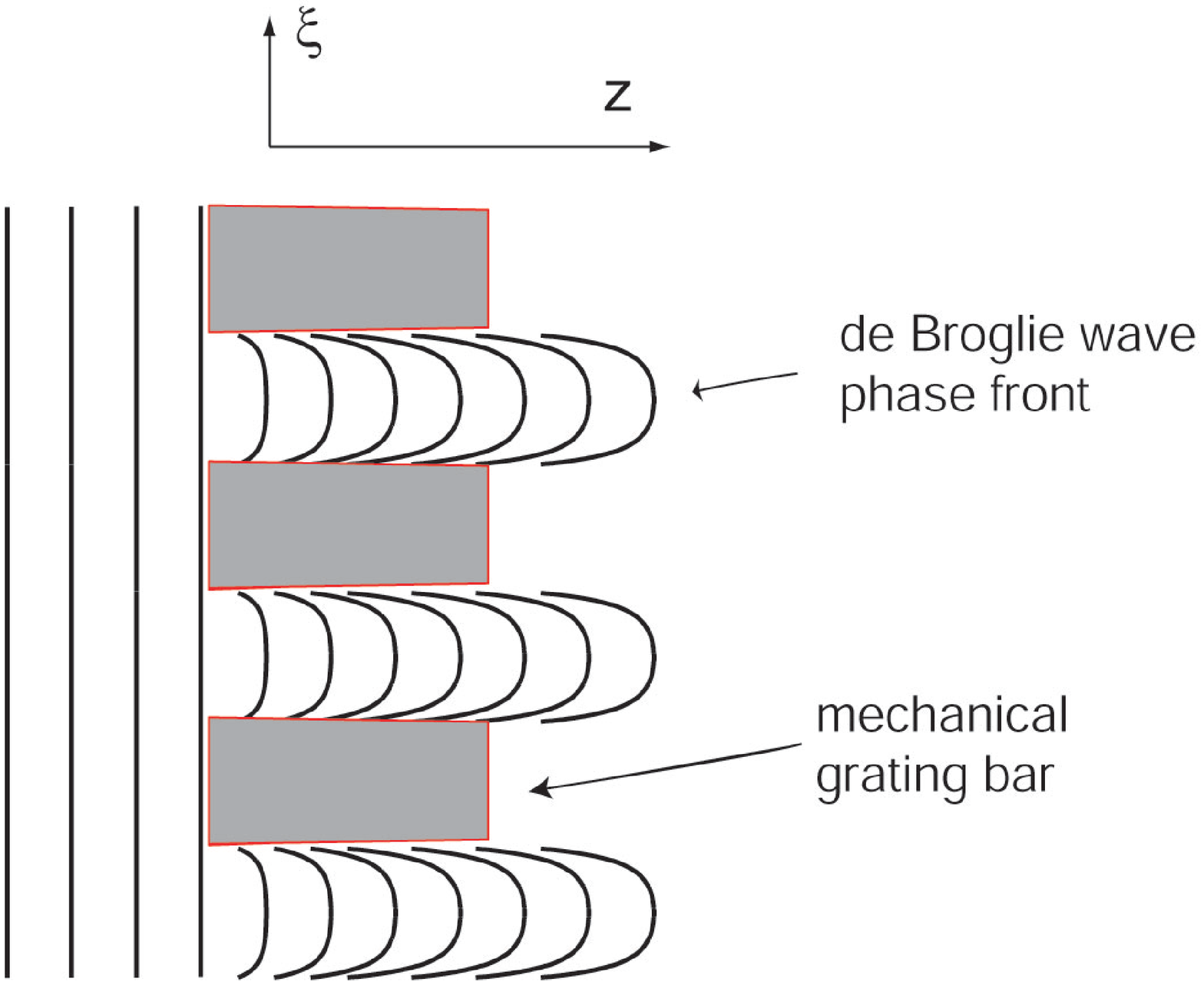}
\caption{\label{wavefronts} Modified wave fronts with $\phi(\xi)$
due to $V(\xi,z)$ calculated by Eqs. \ref{eq:V(r)}  and
\ref{eq:phi(xi)}.}
\end{minipage}
\end{figure}

To begin, we approximate the potential in each channel by a sum of
van der Waals potentials for an atom and two infinite planes,
\begin{equation} V(r) = -C_3\left(\frac{1}{r_1^3} +
\frac{1}{r_2^3}\label{eq:V(r)}\right)
\end{equation} where $r_{1}$ and $r_2$ are the distances along the
normals to the walls of the channel, and $r$ will be defined in
terms of the coordinates $\xi$ and $z$. Because of the trapezoidal
shape of the bars, the channel walls are not parallel. We also
considered the potential due to a sum of uncorrelated atom-atom
interactions between a beam atom and all the atoms that compose the
grating. The potential energy landscape calculated this way is shown
in Figure \ref{3Dpotential} and the marginal validity of this
approach will be discussed later.

The phase shift for atom waves transmitted through the channel can
be computed in the WKB approximation,
\begin{equation} \phi(\xi) = \int \sqrt{\frac{2m}{\hbar^2}[E-V(\xi,z)]}
dz \label{eq:phi(xi)}
\end{equation} where $m$ is the atomic mass, $E$ is the total
atomic energy, and $V(\xi,z)$ is the van der Waals potential energy.
The coordinate axes $\xi$ and $z$ are defined in Figures
\ref{3Dpotential} and \ref{wavefronts}.  Curved wave fronts due to
$\phi(\xi)$ are shown in Figure \ref{wavefronts} given a plane wave
incident on the gratings. With this model of the grating windows as
phase masks, the wave function after the grating has an additional
phase factor that depends on transverse position within each window
given by
\begin{equation} T(\xi) = \mbox{rect}\left(\frac{\xi}{w}\right)e^{i\phi(\xi)}, \end{equation}
where $T()$ is the transmission function, $\xi$ is measured from
the center of each window and rect() describes the absorption from
the grating bars. Then the wave function at the detector plane is
a sum of diffraction orders
\begin{equation} \psi(x) = \sum_n |A_n| e^{i\Phi_n} L(x-x_n)
\end{equation} with
\begin{equation} |A_n|  e^{i\Phi_n} = \int T(\xi) d \xi =
\int_{-w/2}^{w/2} e^{i\phi(\xi)+ink_g\xi}  d \xi
\label{eq:An=integ...}
\end{equation} where $k_g=2\pi/d$ is the grating wavenumber, $w$
is the size of each window between the grating bars, and $L(x)$ is
the line shape of the atom beam in the detector plane, and $x_n$
is the displacement of the $n$th diffraction order given by $x_n =
z_{det}n\lambda_{dB}/d$. In our experiment $z_{det} = 2.4$ m,
\mbox{$d = 100$ nm,} and $\lambda_{dB}$ can range from 270 pm  \
(for 600 m/s Na atoms) to 54 pm  \ (for 3000 m/s Na atoms).  More
formal derivations of $\psi(x)$ are given in \cite{CRP04,PCS05}
from our group and in \cite{GST99,BRU02} from the Toennies group
which uses a slightly different description of the atom optics
theory.


\begin{figure}[h]
\begin{minipage}{18pc}
\includegraphics[width=14pc]{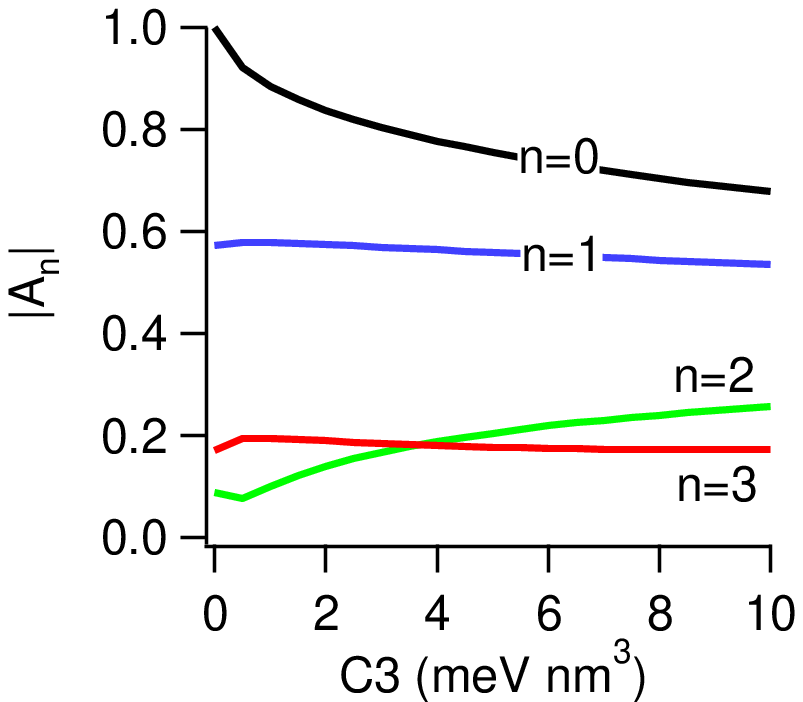}
\caption{\label{an}Diffraction amplitudes $|A_n|$ depend on $C_3$.
The prediction includes fixed parameters ($v$ = 2000 m/s, $w = 55$
nm, $d = 100$ nm, $t=150$ nm, and $\alpha = 5^o$).}
\end{minipage}\hspace{2pc}%
\begin{minipage}{18pc}
\includegraphics[width=14pc]{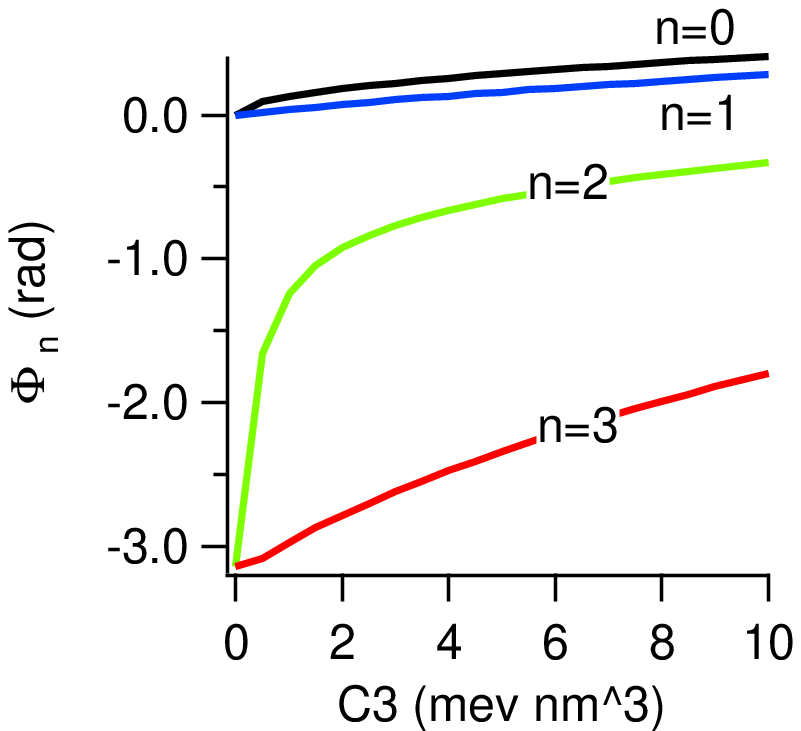}
\caption{\label{fn}Phases $\Phi_n$ for each diffraction order depend
on $C_3$.  The prediction includes fixed parameters ($v$ = 2000 m/s,
$w = 55$ nm, $d = 100$ nm, $t=150$ nm,   and $\alpha = 5^o$).}
\end{minipage}
\end{figure}

The phase ($\Phi_n$) and amplitude ($|A_n|$) of each diffraction
order are both functions of the potential strength ($C_3$), the
atom velocity ($v$), grating window size ($w$), grating period
($d$), grating thickness ($t$), and grating bar wedge angle
($\alpha$). For typical values ($v$ = 2000 m/s, $w = 55$ nm, $d =
100$ nm, $t=150$ nm,   and $\alpha = 5^o$) theoretical plots of
$|A_n(C_3)|$ and $\Phi_n(C_3)$ are shown in Figures \ref{an} and
\ref{fn}.

\section{Atom and Electron Diffraction Experiments}

\begin{figure}[h]
\begin{minipage}{18pc}
\includegraphics[width=18pc]{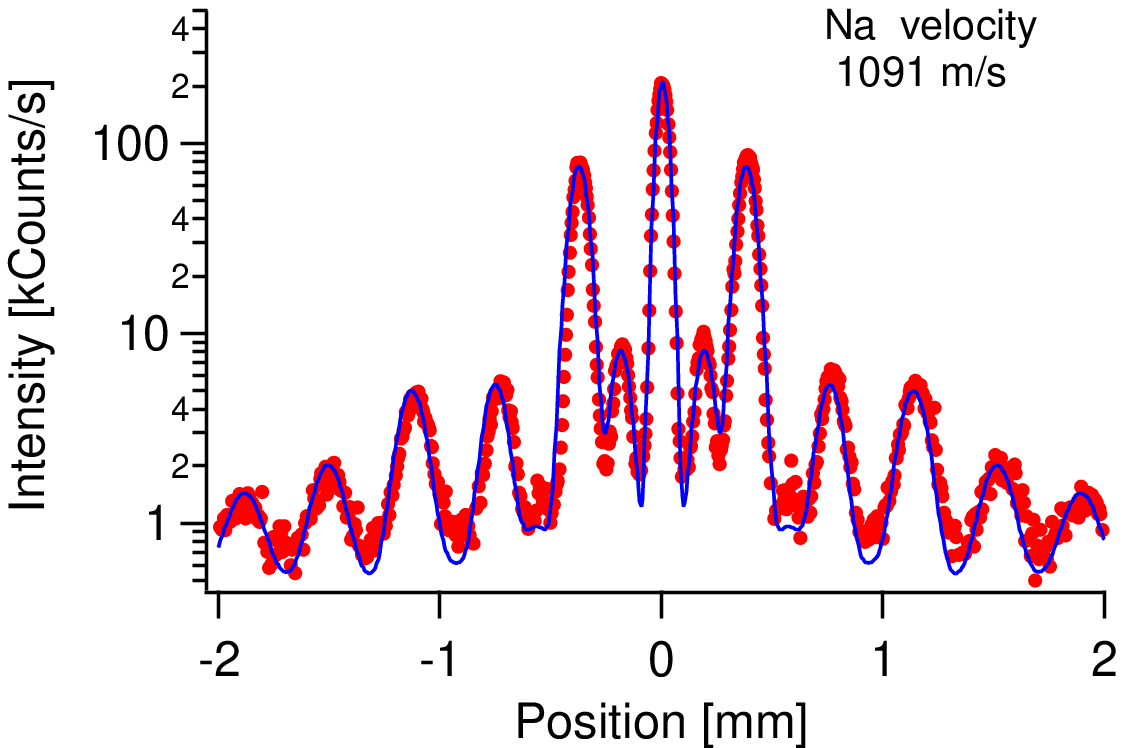}
\caption{\label{dif1000}Sodium atom (and molecule) diffraction
data using a supersonic beam seeded with Ar carrier gas to provide
mean velocity of 1091 m/s.}
\end{minipage}\hspace{2pc}%
\begin{minipage}{18pc}
\includegraphics[width=18pc]{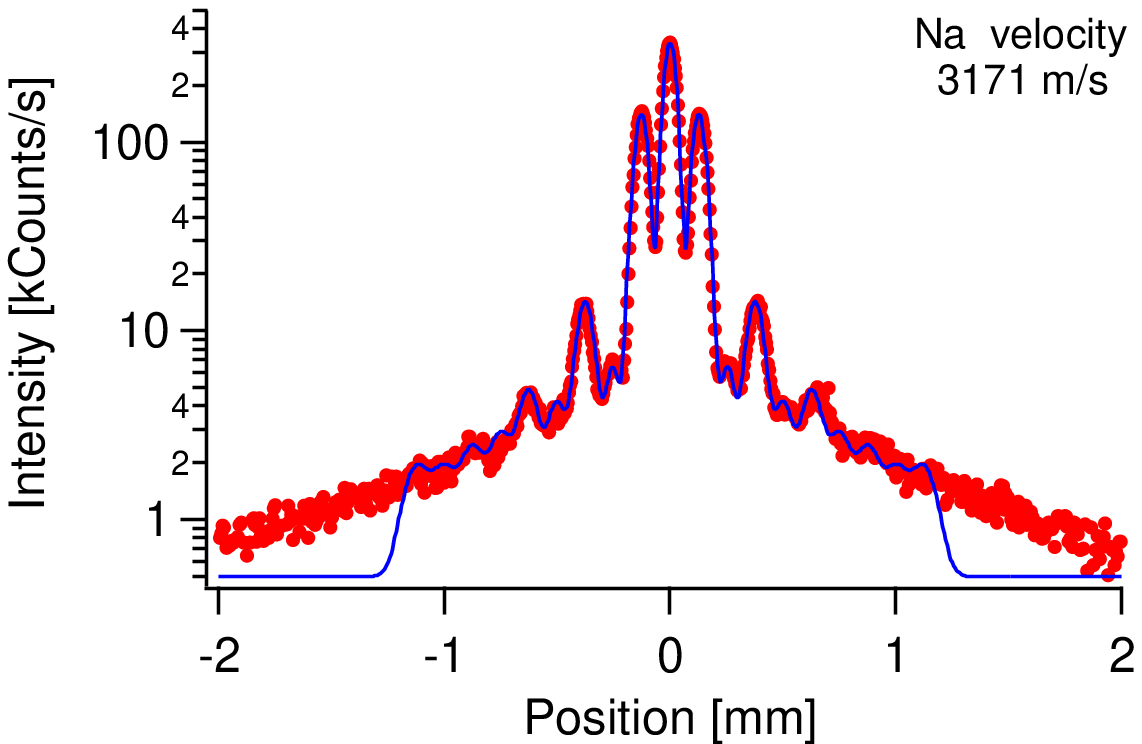}
\caption{\label{dif3000}Sodium atom (and molecule) diffraction
data using a supersonic beam seeded with He carrier gas to provide
mean velocity of 3171 m/s.}
\end{minipage}
\end{figure}

Diffraction data displayed in Figure \ref{dif1000} permit us to
measure the intensities $|A_n|^2$ and the mean atom beam velocity,
as well as the velocity distribution. When combined with SEM
measurements of the grating geometry, a single diffraction pattern
is sufficient to determine $C_3$, because the values of $|A_n|^2$
depend on $C_3$ as shown in Eqs. \ref{eq:V(r)}, \ref{eq:phi(xi)}
and \ref{eq:An=integ...}.

Ideally, we would like to manipulate $C_3$ and verify
experimentally that this changes $|A_n|$. Instead, we studied how
each $|A_n|$ for $n$=(0 to 5) depends on atomic velocity. For
example,  Figures \ref{dif1000} and \ref{dif3000} show different
diffraction intensities ($|A_n|^2$) for two different velocity Na
beams. As expected, the diffraction angle changes too, and this
permits us to measure $v$. To first order in $V(r)/E$ the
parameters $C_3 t / v$ are grouped together in $\phi(\xi)$ and
therefore $|A_n|$ are affected in a similar way by changing $v$ or
$C_3$. This experiment is described in \cite{PCS05}.

We emphasize here that if $C_3=0$, the nano structure would be
purely an absorbing grating, and the $|A_n|$ would not change with
atom velocity.  It is the van der Waals interaction that gives the
gratings a complex transmission function, and makes $|A_n|$ depend
on incident atom velocity. With least squared fits to $|A_n|$ we
determined $C_3=2.7\pm0.8$ meVnm$^3$. We also verified using
$\chi^2$that the $1/r^3$ form of the potential gives the best fit
to the data.

\begin{figure}[h]
\begin{minipage}{18pc}
\includegraphics[width=18pc]{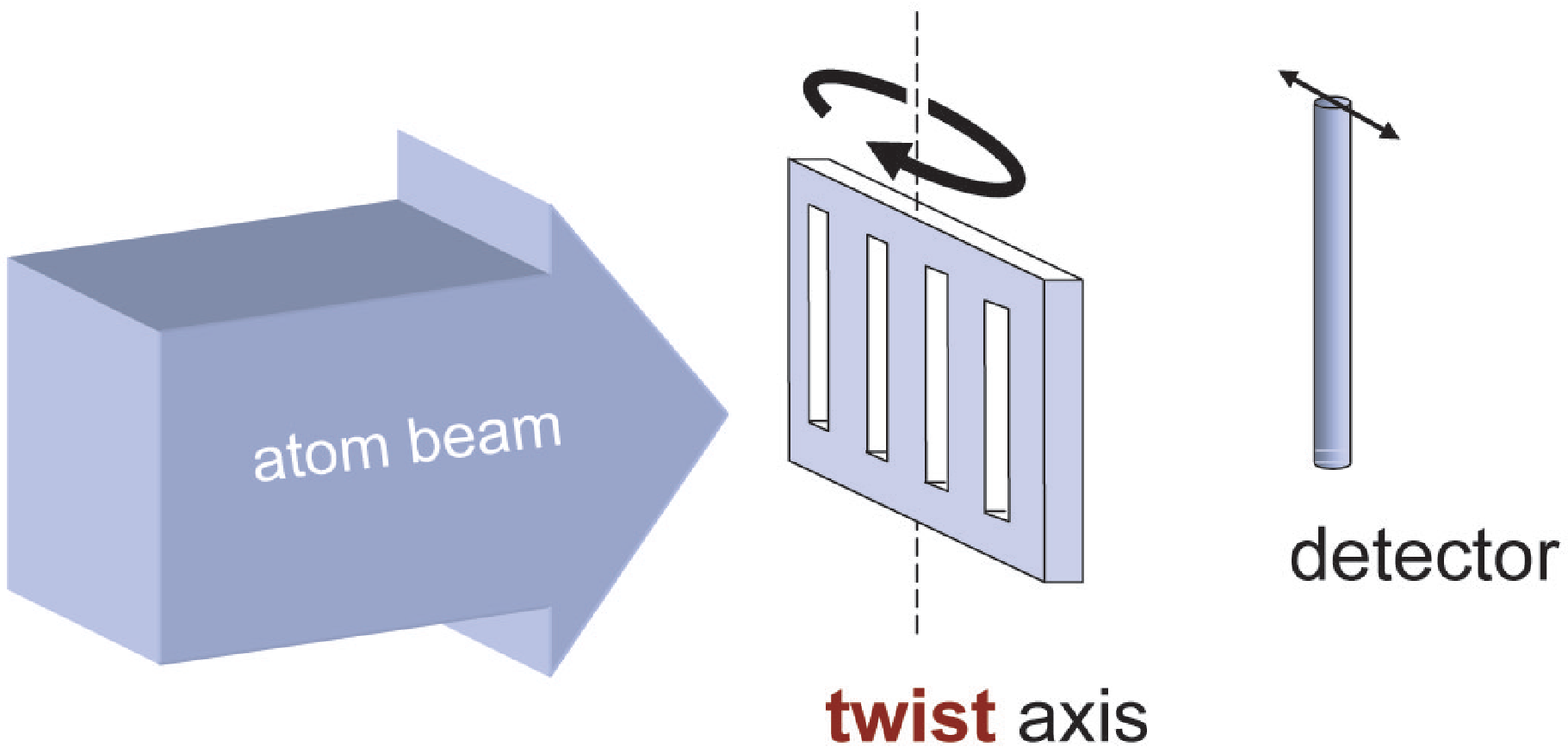}
\caption{\label{rot}Geometry of the diffraction experiments.  In
\cite{PCS05} the beam velocity is changed.  In \cite{CRP04} the
grating is rotated.}
\end{minipage}\hspace{2pc}%
\begin{minipage}{18pc}
\includegraphics[width=13pc]{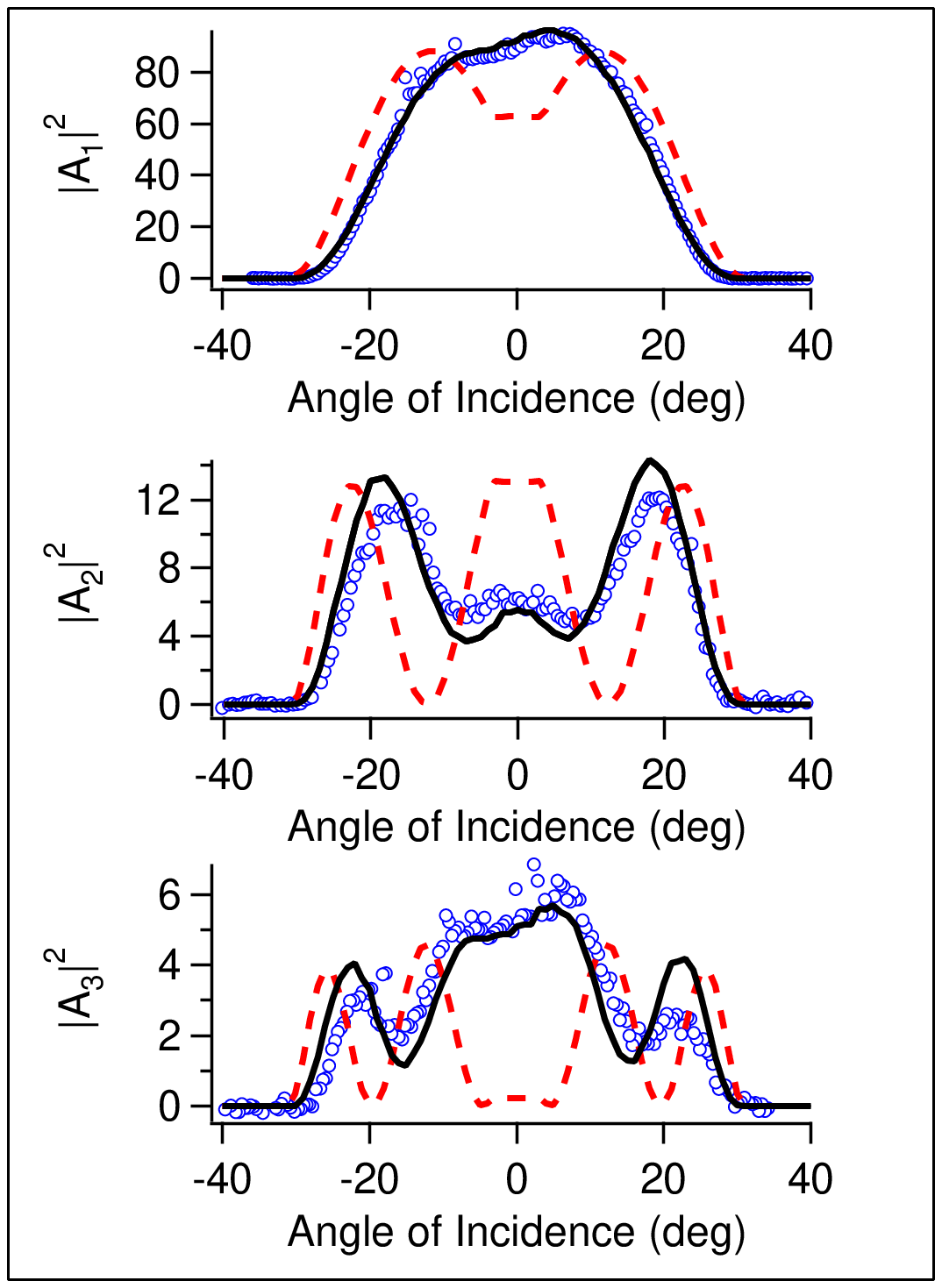}
\caption{\label{rot1}The intensity in the first three diffraction
orders ($|A_1|^2$, $|A_2|^2$, and $|A_3|^2$) as a function of the
angle of incidence.}
\end{minipage}
\end{figure}

In our second experiment we observed how the intensities $|A_n|^2$
vary when the grating is rotated as shown in Figures \ref{rot} and
\ref{rot1} and described in \cite{CRP04}. Due to the thickness of
the grating bars the angle of incidence affects the projected open
fraction.  If $C_3=0$ then twisting the grating would cause
missing diffraction orders when the projected open fraction is
$\frac{1}{2}$, or $\frac{1}{3}$ as shown with dashed lines in
Figure \ref{rot1}. However, a better fit to the data is obtained
using the theory for $|A_n|$ described in \cite{CRP04} (similar to
Eq. \ref{eq:An=integ...}) with $C_3=5$ meVnm$^3$. This produces
the solid lines in Figure \ref{rot1}.
We also observe asymmetric diffraction patterns, i.e.
\emph{blazed} diffraction, in this experiment, and our model
reproduces this result. Asymmetric phase profiles $\phi(\xi)$ for
de Broglie waves transmitted through each window are provided by a
combination of three ingredients: non-normal incidence,
trapezoidal bar shape, and non-zero $C_3$. We observe even more
asymmetric diffraction patterns with electron beams as discussed
next.

\begin{figure}[h]
\begin{minipage}[h]{20pc}
\includegraphics[width=14pc]{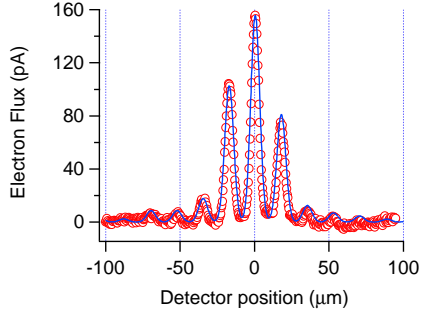}\hspace{2pc}
\end{minipage}
\begin{minipage}[h]{18pc}\caption{\label{edif}Electron diffraction
from a beam transmitted through a nanostructure grating.  The
theoretical curve includes the effect of the image-charge on the
diffraction intensities $|A_n|^2$.}
\end{minipage}
\end{figure}

We found that electron diffraction with nanostructure gratings is
also affected by the surfaces as far as 20 nm away, similar to the
way van der Waals interactions modify atom diffraction. To keep the
grating bars from getting charged by the electron beam we coated
them with 1 nm of gold. Near this gold-coated silicon nitride
surface, electrons have an electrostatic potential energy that we
calculate from the method of images to be
\begin{equation}V(r)=-e^2\left(\frac{1}{2r_1}+\frac{1}{2r_2}\right)
 \left(\frac{\epsilon-1}{\epsilon+1} \right) \label{eq:image}
\end{equation} in Gaussian units, where $\epsilon$ is the ratio of
the dielectric permittivity compared to that of free space and $e$
is the charge of the electron and as before, $r_1$ is the distance
to one wall. Using this image-charge potential we calculated the
phase shift $\phi(\xi)$ and thus the diffraction amplitudes
$|A_n|$ for electrons.  Figure \ref{edif} shows electron
diffraction data and a theoretical diffraction pattern based on
Equations 2, 5, and \ref{eq:image}, using best fit parameters
$\epsilon=4$, and an angle of incidence $5$ degrees. Note that if
the image-charge effect were not included then the model for
$|A_n|^2$ in Figure \ref{edif} would be symmetric about the zeroth
order.

Twenty nm away from this surface the potential for electrons is
0.02 eV as compared to the atom-wall potential of 0.4 $\mu$eV.
This is 50,000 times larger for electrons than for sodium atoms.
However, the phase shift $\phi(\xi)$ near $\xi=20$ nm is similar
for electrons and atoms because the velocity of 500 eV electrons
(13,000 km/s) is four orders of magnitude larger than that for
0.12 eV sodium atoms (1 km/s).

\section{Atom Interferometer Experiments}

\begin{figure}[h]
\includegraphics[width=18pc]{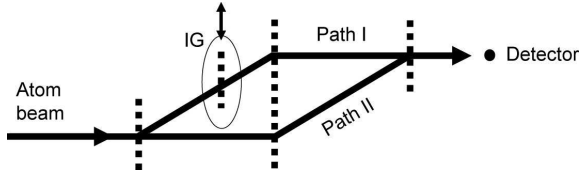}\hspace{2pc}
\begin{minipage}[b]{18pc}\caption{\label{ifm} Top view of the atom
interferometer.  Three gratings make the interferometer and a
removable interaction grating (IG) is used in \cite{PEC05} to
study the phase shift $\Phi_0$ caused by the van der Waals
interaction.}
\end{minipage}
\end{figure}

\begin{figure}[h]
\begin{minipage}{18pc}
\includegraphics[width=16pc]{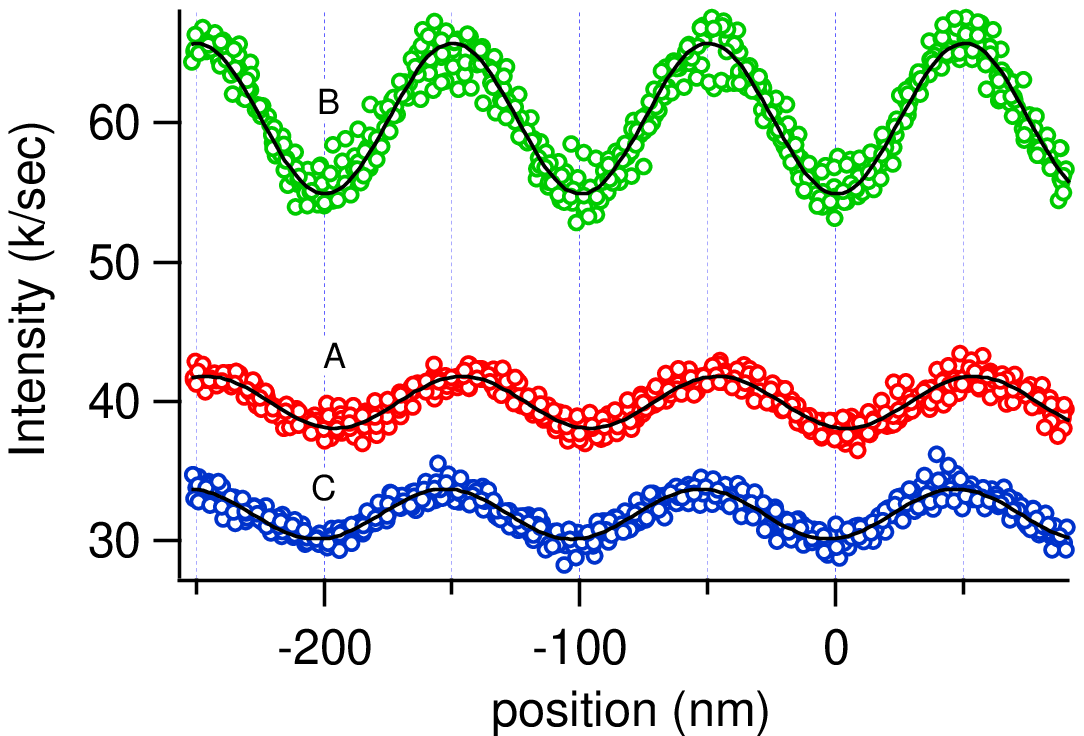}
\caption{\label{fringes}Atom interference fringe data for three
cases: (A) interaction grating (IG) located in path \textsf{I}, (B)
no IG,  (C)  IG in path \textsf{II} of the atom interferometer.}
\end{minipage}\hspace{2pc}%
\begin{minipage}{18pc}
\includegraphics[width=16pc]{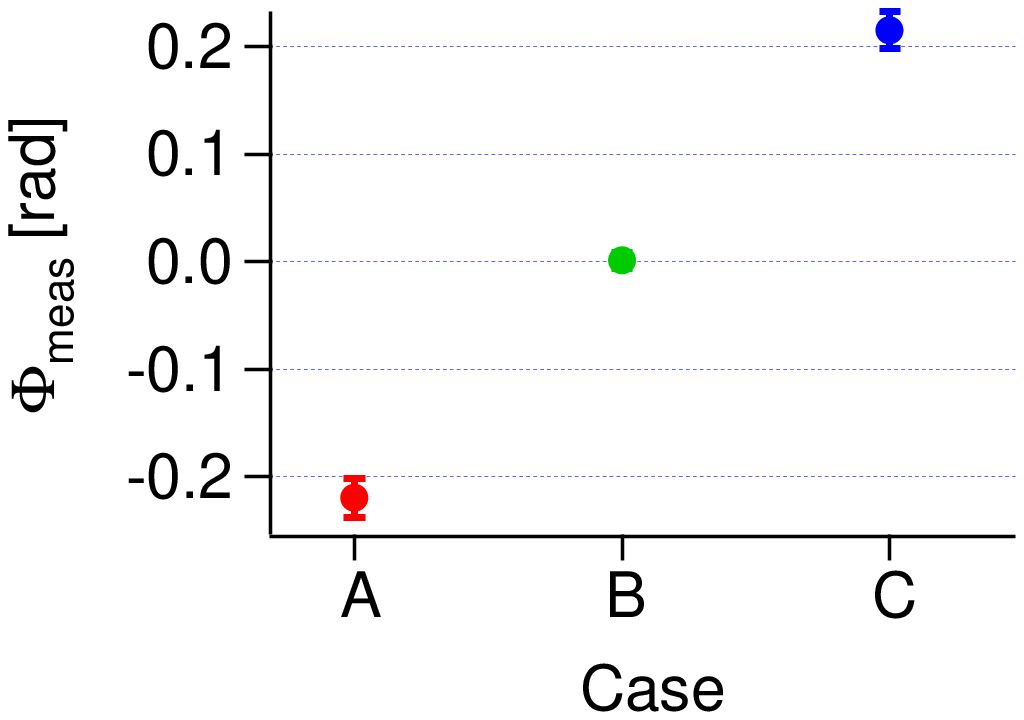}
\caption{\label{label}Best fit phase and statistical error bars for
the three cases: (A) IG in \mbox{path \textsf{I}}, (B) IG removed,
(C) IG in path \textsf{II} of the interferometer.}
\end{minipage}
\end{figure}

In addition to the diffraction amplitudes $|A_n|$, there are
diffraction phases $\Phi_n$ in each order. We measured the phase
shift in the zeroth order ($\Phi_0$) due to transmission through a
removable interaction grating (IG) with an interferometer
\cite{PEC05}. The geometry of this experiment is shown in Figure
\ref{ifm}, and permits us to study the interference fringes as a
function of the IG position. Interference fringe data are shown in
Figure \ref{fringes} for the cases of the IG obscuring either path
\textsf{I}, path \textsf{II}, or neither. The directly measured
phase shift, $\Phi_{meas} = 0.22 \pm 0.02$ radians, is consistent
with $C_3 = 4\pm1$ meVnm$^3$. A more detailed discussion of how to
determine $\Phi_0$ and $C_3$ from the measured phase shift is
given in a separate entry of the CAMS conference proceedings by
J.D. Perreault and in reference \cite{PEC05}.

We also measured the phase difference $\Phi_2 -\Phi_1= 0.6 \pm
$0.4 radians, by comparing the relative phases of four separate
interferometers formed by adjacent pairs of the $n=(-2,-1,0,1,2)$
diffraction orders of the first interferometer grating. This phase
difference can be compared to that predicted in Figure \ref{fn}
and is consistent with $C_3 = 4$ meVnm$^3$.

\section{Discussion}

\begin{figure}[h]
\begin{minipage}[h]{20pc}
\includegraphics[width=18pc]{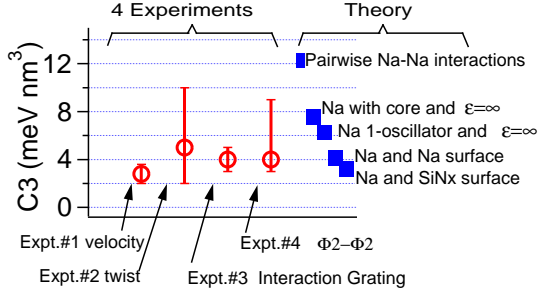}\hspace{2pc}
\end{minipage}
\begin{minipage}[h]{18pc}\caption{\label{comparison}Comparison of
$C_3$ values determined by the four experiments presented here (red
circles).  Theoretical predictions based on different descriptions
of the atom and the surface are shown (blue squares).}
\end{minipage}
\end{figure}

To compare our measurements with theory, we reviewed calculations of
$C_3$ for sodium atoms and various surfaces as shown in Figure
\ref{comparison}. The Lifshitz formula \cite{Lif56} for $C_3$ is
\begin{equation} C_3 = \frac{\hbar}{4\pi}\int_0^{\infty}
\alpha(i \omega) \frac{\epsilon(i\omega)-1} {\epsilon(i\omega)+1}d
\omega \label{eq:lifshitz} \end{equation} where $\alpha(i \omega)$
is the polarizability of the atom and $\epsilon(iw)$ is the
permittivity of the surface. For a perfect conductor
($\epsilon$=$\infty$) and sodium atoms, Derevianko $et$ $al.$
\cite{DJS99} calculated $C_3= 7.60$ meVnm$^3$ and noted that $16\%$
of this value is due to the core electrons. A single Lorentz
oscillator model for an atom with no damping gives the
polarizability: \begin{equation} \alpha(i\omega) =  \frac
{\alpha(0)} {1+ (\frac{\omega}{\omega_0})^2}.
\label{eq:lorentz}\end{equation} For sodium atoms $\alpha(0)=0.0241$
nm$^3$ \cite{ESC95} and $\omega_0 = 2\pi c/(590 \ $nm$)$.  Combining
this with ($\epsilon$=$\infty$) in Eq.~\ref{eq:lifshitz} gives $C_3=
\hbar\omega_0 \alpha(0)/8 = 6.3 \ $meVnm$^3$. This agrees with the
non-retarded limit in \cite{MDB97}.

For a metal surface, the Drude model describes $\epsilon(i\omega)$
with the plasma frequency and damping:
\begin{equation} \epsilon(i \omega) = 1 + \frac{\omega_p^2}
{\omega(\omega + \gamma)}.\label{eq:bulk sodium epsilon}
\end{equation} For sodium metal,
$\hbar \omega_p = 5.8  $ eV and $\hbar \gamma = 23 $meV, 
resulting in $C_3 = 4.1 $ meV nm$^3$ for a sodium atom and a bulk
sodium surface.  For an insulating surface Bruhl $et$ $al$
\cite{BRU02} used a model with
\begin{equation} \epsilon(i \omega) = \frac{\omega^2 +
(1+g_0)\omega_0^2} {\omega^2 + (1-g_0)\omega_0^2} \end{equation} and
$\hbar \omega_0 \equiv E_s = 13$ eV and $g_0 = 0.588$ for silicon
nitride. Using this expression and the one-oscillator model for
sodium atoms yields $C_3 = 3.25 \ $ meVnm$^3$.

The pairwise sum of atom-atom interactions for sodium atoms near
bulk sodium metal gives
\begin{equation} V(x) = -N\int_x^{\infty} \int_{-\infty}^{\infty}
\int_{-\infty}^{\infty} \frac{C_6}{r^6} dy' dz' dx' = -\frac{\pi N
C_6}{6x^3}\end{equation} where $x$ is the atom-surface distance.
Using the London result for $C_6 = (3/4)\hbar\omega_0 \alpha(0)^2$
\cite{MIL94} and the number density of bulk sodium for $N$ gives a
value for $C_3$ = 12.3 meVnm$^3$ (also, if $N$ is replaced by
$\alpha(0)^{-1}$, this calculation gives $C_3 = \pi \hbar \omega_0
\alpha(0)/8 = 19.6$meVnm$^3$). The value using the pairwise sum is
three times larger than the value (4.1 meVnm$^3$) obtained for the
same atom-surface system using Eqs.\ref{eq:lifshitz},
\ref{eq:lorentz}, and \ref{eq:bulk sodium epsilon}.  The different
values obtained with these two approaches demonstrate the
non-additivity of the van der Waals potential. Our measurements of
$C_3$ are all much closer to the Lifshitz result.

\begin{figure}[h]
\begin{minipage}[h]{20pc}
\includegraphics[width=11pc]{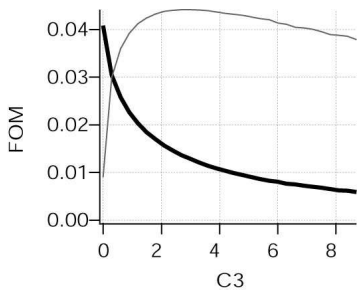}\hspace{2pc}
\end{minipage}
\begin{minipage}[h]{20pc}\caption{\label{fom}  The FOM depends on $C_3$. The
thick curve is for windows ($w_1$=56, $w_2$=50, $w_2$=37 nm) that
maximize the FOM if $C_3=0$. The thin curve is for windows
($w_1$=93, $w_2$=88, $w_2$=37 nm) that maximize the FOM for $C_3 =
3$ meVnm$^3$.  These calculations are for Na atoms with $v=100$
m/s.}
\end{minipage}
\end{figure}

One example of how van der Waals forces influence atom optics with
nanotechnology is that the figure of merit for an interferometer
discussed in \cite{BER97} (FOM =
$\mbox{contrast}\times\sqrt{I/I_{inc}})$ depends on $C_3$. Here we
show that to maximize the FOM, different open fractions need to be
chosen if $C_3 \neq 0$, especially for slower atom beams as shown
in Figure \ref{fom}.



\section{Conclusion}
We used nanostructure gratings with 50-nm wide channels between
free-standing bars in four experiments to measure the strength of
atom-surface interactions. We detected phase shifts and intensity
changes for atom beams that we attribute to van der Waals
interactions with the walls of a nanostructure grating. A model
based on complex transmission functions for de Broglie wave optics
can explain both atom and electron diffraction patterns.

\ack This research was supported by a Research Innovation Award
from the Research Corporation and by National Science Foundation
Grants No.0354947 and ECS-0404350. We thank Tim Savas for making
the gratings and Ben McMorran for analyzing electron diffraction.

\end{document}